\def\be{\begin{equation}}
\def\ee{\end{equation}}
\def\bea{\begin{eqarray}}
\def\eea{\end{eqarray}}

\documentclass{ws-p8-50x6-00}

\begin{document}

\title{Electroweak properties of baryons in a covariant chiral quark model}

\author{S.~Boffi, M.~Radici, L.~Glozman}

\address{Dipartimento di Fisica Nucleare e Teorica,
Universit\`a di Pavia \\ and Istituto Nazionale di Fisica Nucleare, 
Sezione di Pavia, I-27100 Pavia, Italy}

\author{R.F.~Wagenbrunn, W.~Plessas}

\address{Institut f\"ur Theoretische Physik, Universit\"at Graz, \\
Universit\"atsplatz 5, A-8010 Graz, Austria}

\author{W.~Klink}

\address{Department of Physics and Astronomy, University of Iowa, \\
Iowa City, IA 52242, USA}


\maketitle

\abstracts{The proton and neutron electromagnetic form factors and the
nucleon axial form factor have been calculated in the Goldstone-boson exchange
constituent-quark model within the point-form approach to relativistic quantum
mechanics. The results, obtained without any
adjustable parameter nor quark form factors, are, due to the dramatic effects
of the boost required by the covariant treatment, 
in striking agreement with the data.
}

Constituent quark models (CQM's) provide a useful framework for quantitative
calculations of hadron properties in a regime where quantum chromodynamics (QCD)
cannot be solved perturbatively. Constituent quarks are effective degrees of
freedom described in terms of an Hamiltonian reflecting basic symmetries of QCD.
Thus the formalism relies on
quantum mechanics with a finite number of degrees of freedom rather than on
quantum field theory. However, due to the large value of their kinetic energy
constituent quarks are relativistic quasi-particles requiring a relativistic
quantum mechanical formulation in terms of unitary representations of the
Poincar\'e group. 

Technically, the problem is solved by looking at one of the (unitarily
equivalent) forms that are possible when defining the (kinematic) stability
subgroup~\cite{Leutwyler:1978}. Here we adopt the point form that has recently 
attracted some interest in connection with the electromagnetic properties of
hadrons~\cite{Klink:1998hb}. In fact, this form has some advantages. First,
the four-momentum operators $P^\mu$ containing all the dynamics commute with
each other and can be simultaneously diagonalized. Since the Lorentz generators do
not contain any interaction terms, the theory is manifestly covariant. Second,
the electromagnetic current operator $J^\mu(x)$ can be written in such a way
that it transforms as an irreducible tensor operator under the strongly
interacting
Poincar\'e group. Thus the nucleon charge and
magnetic form factors can be calculated as reduced matrix elements of such an
irreducible tensor operator in the Breit frame.

Here results are presented as a progress report of a more comprehensive
programme dealing with
electroweak properties of baryons studied within the CQM discussed in
ref.~\cite{Glozman:1998ag} and based on
Goldstone-boson-exchange (GBE) dynamics. This type of CQM assumes a pairwise
linear confinement potential, as suggested
by lattice QCD, with a strength according to the string tension of QCD. The
quark-quark interaction is derived from the exchange of pseudoscalar bosons
producing the (flavour dependent)
hyperfine interaction; in the model only the spin-spin component is utilized,
which 
phenomenologically appears to be the most important in the hyperfine splitting
of the baryon spectra. The current operator is a single-particle current
operator for point-like constituent quarks. This approach corresponds
to a relativistic impulse approximation but specifically in point form. It is
called point-form spectator approximation (PFSA).

\begin{figure}[t]
\vspace*{-0.3cm}
$
\begin{array}{c}
\hspace*{1.5cm}\psfig{file=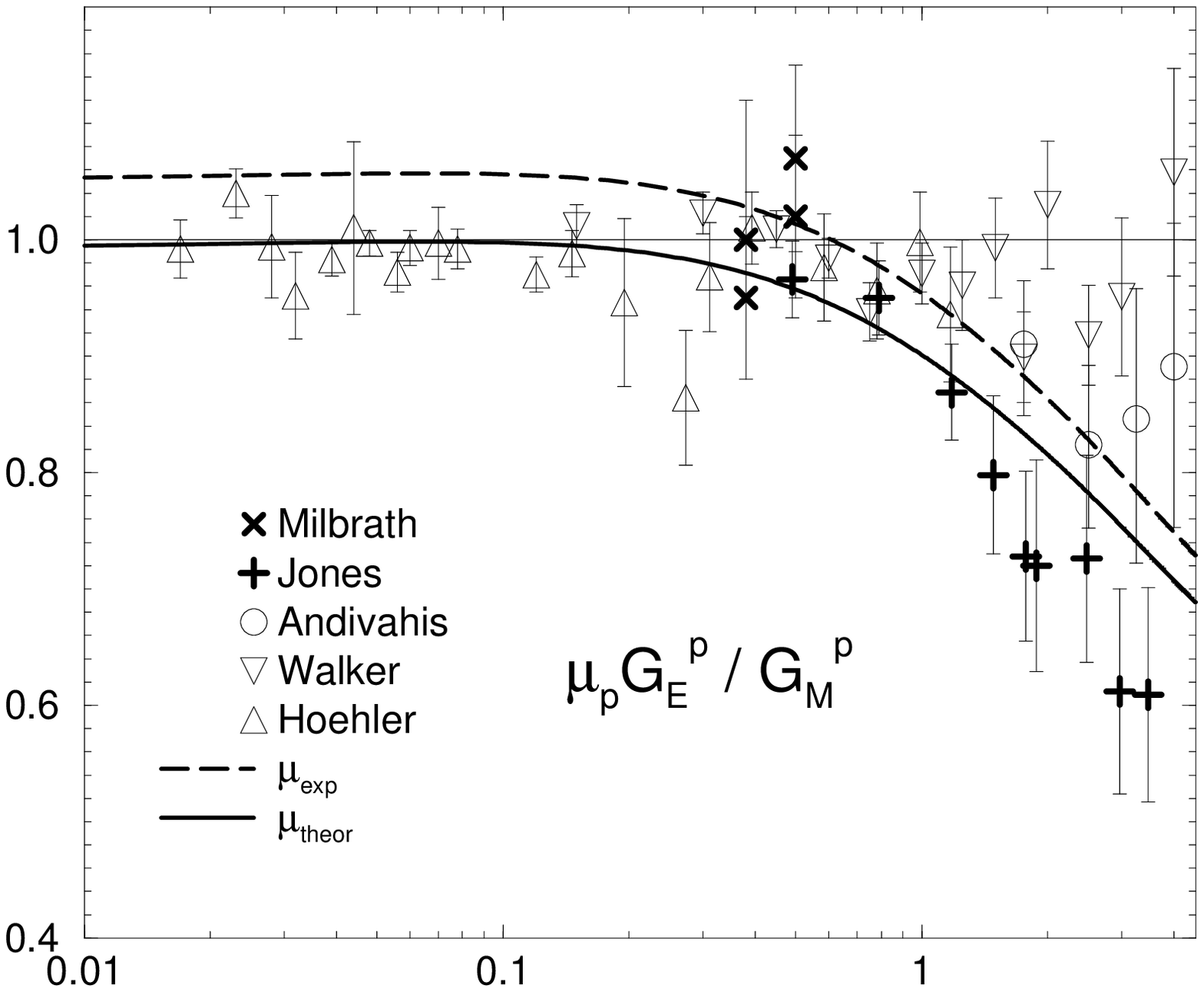,width=19pc}\\[-0.9cm]
\hspace*{1.5cm}\psfig{file=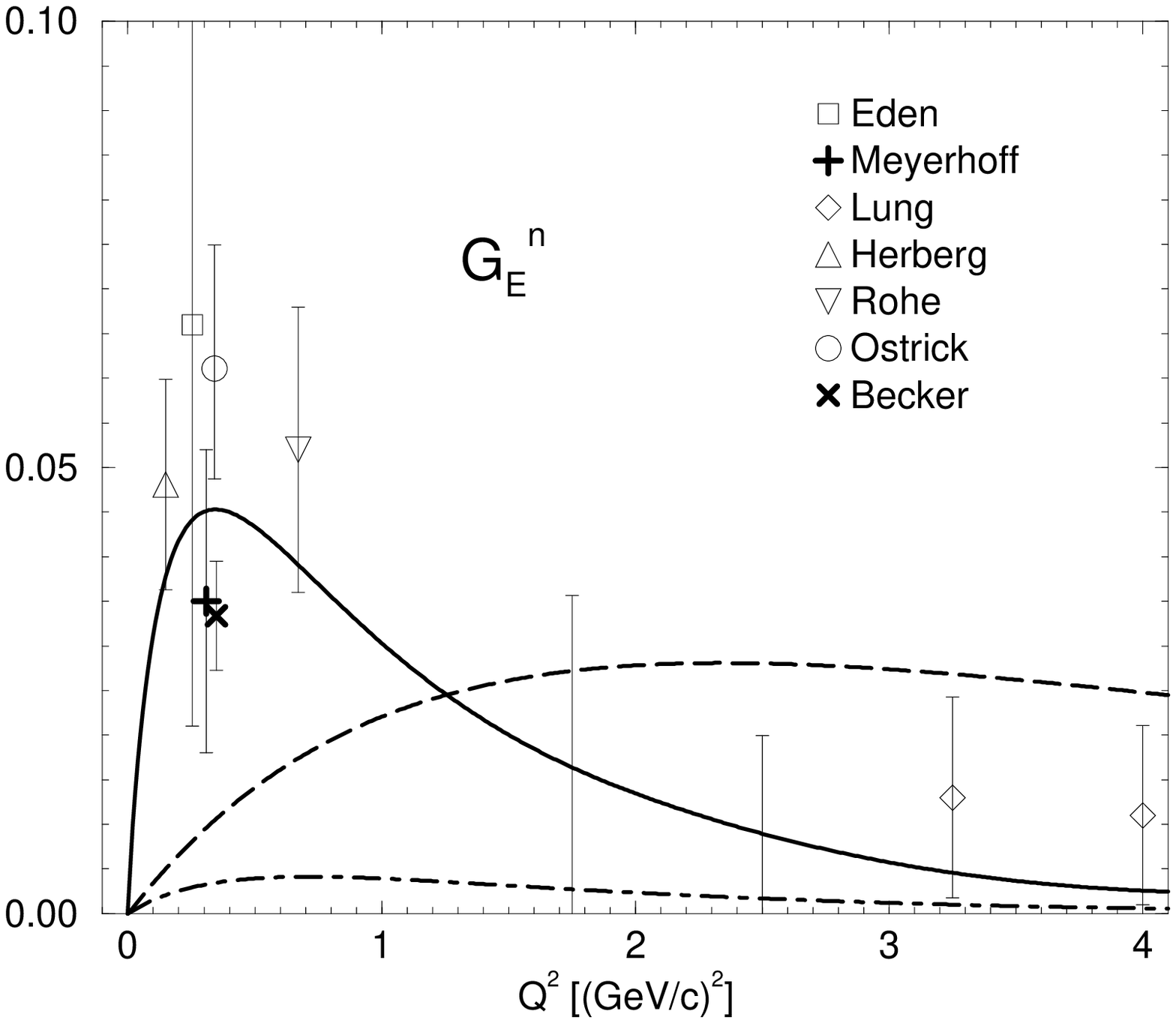,width=19pc}\\[-0.4cm]
\end{array}
$
\caption{Top: the ratio of the proton electric to magnetic form factor. 
Solid (dashed) line obtained with the theoretical (experimental) value of the
magnetic moment $\mu$. Bottom: the neutron electric form factor. Solid, dashed
and dot-dashed lines as predicted by the model in PFSA, the nonrelativistic
approximation and the model with the confinement interaction only, respectively.
\label{fig:ff} }
\vspace{-0.5cm}
\end{figure}

In fig.~\ref{fig:ff} the ratio $G_E/G_M$ of the proton electric to magnetic form
factor is shown
together with the recent TJNAF data~\cite{Jones:1999rz}. In fig.~\ref{fig:ff} 
the prediction of the model for the neutron charge form factor is also shown.
The solid (dashed) curve is obtained when the theoretical (experimental) value
of the proton magnetic moment is used. The difference between the two curves is
due to the fact that the calculated proton and neutron
charge radii turn out to be in very good agreement with experiment ($r^2_p =
0.75$ fm$^2$, $r^2_n = -0.12$ fm$^2$) while the magnetic moments are slightly
underestimated ($\mu_p = 2.64$ n.m., $\mu_n = -1.67$ n.m.). However, 
one observes that a very good description of both
the proton and neutron e.m. structure is achieved~\cite{Covariant}. It is
remarkable that no further ingredients beyond the quark model wave functions
(such as, e.g., constituent quark form factors) have been employed. What is
importantant is that only
relativistic boost effects are properly included in point-form relativistic
quantum mechanics. 

For comparison, results for the neutron charge form factor are also shown  when
calculated in nonrelativistic impulse approximation, i.e. with the
standard nonrelativistic form of the current operator and no Lorentz boosts
applied to the nucleon wave functions. Also the case with the confinement
potential only has been considered in order to appreciate the role of
mixed-symmetry components in the wave functions that are absent without the
hyperfine interaction.

A similar approach can be used to study the axial current~\cite{Axial}. 
According to the PCAC hypothesis this current is not conserved. However,
one can always split it into conserved and nonconserved parts
with the conserved part containing the axial form factor $G_A$ only. Therefore
one can calculate $G_A$ in the same point-form approach
used  for $G_E$ and $G_M$.

\begin{figure}[t]
\vspace*{-0.3cm}
\hspace*{1.5cm}\psfig{file=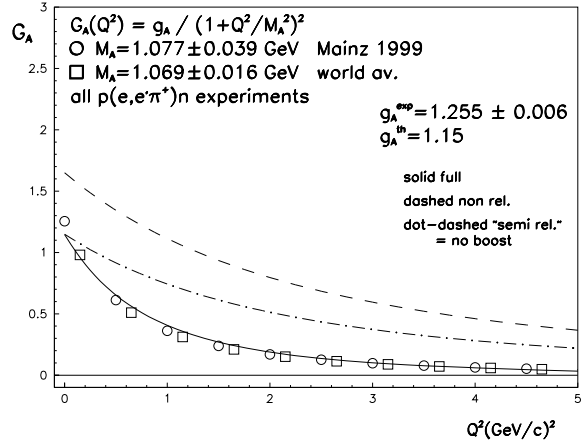,width=20pc}
\vspace{-0.5cm}
\caption{The nucleon axial form factor.}
\label{fig:gamainz}
\vspace{-0.5cm}
\end{figure}

The results are shown in fig.~\ref{fig:gamainz},
where comparison is made with data obtained from charged pion
electroproduction on protons (see ref.~\cite{Liesenfeld} and
references therein). The
nonrelativistic result (dashed line) and the calculation without boost
(dot-dashed line) are also shown. 

The result is quite good at $Q^2\ne 0$, while at $Q^2=0$ the value of the
calculated axial form factor $G_A(0)$ is lower than the value $g_A$
used when fitting the data with a dipole form factor $G_A(Q^2) =
g_A/(1+Q^2/M_A^2)^2$ involving the axial mass $M_A$. This indicates a deviation
of the present results from the usually assumed dipole form.
Correspondingly, the axial radius deduced from the slope at $Q^2=0$ is lower
than the experimental one. With the present model one has
$<r_A^2>^{1/2} = 0.520 \ \hbox{\rm fm}$ to be compared with the experimental
value $<r_A^2>^{1/2} = (0.65\pm 0.07)\ \hbox{\rm fm}$ extracted from neutrino
experiments and $<r_A^2>^{1/2} = (0.635\pm 0.023)\ \hbox{\rm fm}$ extracted from
pion electroproduction~\cite{Liesenfeld}.

\end{document}